\documentclass[preprintnumbers,aps,pra,twocolumn,showpacs]{revtex4}

\usepackage{graphicx}% Include figure files
\usepackage{dcolumn} % Align table columns on decimal point
\usepackage{amsmath}
\usepackage{exscale}
\usepackage{bm}      % bold math
\usepackage{color}
\usepackage{epsfig,psfrag,subfigure,subeqnarray}
\usepackage{bbm}
\usepackage{amsbsy}
\usepackage{amssymb}
\usepackage{enumitem}

\def\lan{\langle}
\def\ran{\rangle}
\def\va{\varepsilon}

\def\vK{{\bf K}}

\def\vp{{\bf p}}

\newcommand{\bd}{\begin{equation}}
\newcommand{\ed}{\end{equation}}
\newcommand{\be}{\begin{equation}}
\newcommand{\ee}{\end{equation}}
\newcommand{\bt}{\begin{split}}
\newcommand{\et}{\end{split}}

\newcommand{\bn}{\begin{align}}
\newcommand{\en}{\end{align}}

\newcommand{\bea}{\begin{eqnarray}}
\newcommand{\eea}{\end{eqnarray}}
\newcommand{\ba}{\begin{array}}
\newcommand{\ea}{\end{array}}
\newcommand{\nn}{\nonumber}

\begin{document}

\title{Mysterious dimensionality effect: the cancellation of the $N$-coboson correlation energy under a BCS-like potential }

\author{ Shiue-Yuan Shiau$^1$, Monique Combescot$^{2}$, and Yia-Chung Chang$^{3,1}$}
\email{yiachang@gate.sinica.edu.tw}
\affiliation{$^1$ Department of Physics, National Cheng Kung University, Tainan, 701 Taiwan}
\affiliation{$^2$Institut des NanoSciences de Paris, Universit\'e Pierre et Marie Curie, CNRS, 4 place Jussieu, 75005 Paris}
\affiliation{$^3$Research Center for Applied Sciences, Academia Sinica, Taipei, 115 Taiwan}

\begin{abstract}
 We use Richardson-Gaudin exact equations to derive the ground-state energy of $N$ composite bosons (cobosons) interacting via a potential which acts between fermion pairs having zero center-of-mass momentum, that is, a potential similar to the reduced BCS potential used in conventional superconductivity. Through a density expansion, we show that while, for 2D systems, the $N$-coboson correlation energy undergoes a surprising cancellation which leaves the interaction part with a $N(N-1)$ dependence only, such a cancellation does not exist in 1D, 3D, and 4D systems --- which corresponds to 2D parabolic traps --- nor when the cobosons interact via a  similar short-range potential but between pairs having an arbitrary center-of-mass momentum. This shows that the previously-found cancellation which exists for the Cooper-pair correlation energy results not only from the very peculiar form of the reduced BCS potential, but also from a quite mysterious dimensionality effect, the density of states for Cooper pairs feeling the BCS potential being essentially constant, as for 2D systems. 

\end{abstract}
\date{\today}

\maketitle

\section{Introduction}

The success of the BCS theory\cite{BCS,BCS2} for conventional superconductors remains fascinating today for its impressive agreement with experimental data, in spite of its extreme simplicity. One of its insightful ingredients is that the attraction between free electron pairs with opposite spins and opposite momenta is enough to capture the physics of superconductivity induced by Cooper pairing. This attractive potential reads 
\be
V_{BCS}=-V\sum_{\vp\vp'} w_{\vp}w_{\vp'} B^\dag_{{\bf0}\vp'}B_{{\bf0}\vp }\label{def:V}
\ee
where $B^\dag_{{\bf0}\vp} \equiv a^\dag_{\vp }b^\dag_{-\vp}$ creates a zero-momentum pair made of an up-spin $(\vp)$ electron and a down-spin $(-\vp)$ electron.
%\be
%V_{BCS}=-V\sum_{\vp\vp'} w_{\vp}w_{\vp'} a^\dag_{\vp' \uparrow }a^\dag_{-\vp' \downarrow}a_{-\vp \downarrow}a_{\vp \uparrow }\label{def:VBCS}
%\ee
 $V$ denotes the potential amplitude, taken as positive and constant, and $w_\vp$ designates the energy layer close to the normal electron Fermi energy $\va_F$, in which a phonon-mediated attraction acts; $w_\vp$ is taken equal to 1 within this energy layer and zero outside. \
 
 %: $w_\vp$ is equal to 1 for $\va_{F_0}\leq \va_\vp \leq  \va_{F_0}+\Omega$ and zero otherwise, with $\Omega$  being of the order of a phonon energy  and $\va_{F_0}=\va_F-\Omega/2$.\

This so-called ``reduced BCS potential" leads to one of the very few exactly solvable many-body problems. Richardson\cite{Richardson1,Richardson2,Richardson3,Richardson4} and Gaudin\cite{Gaudin1976} have independently shown that the ground-state energy of $N$ pairs reads as $E_N=\sum_{i=1}^N R_i$, where the $R_i$'s are solution of the $N$ coupled equations
\be
\frac{1}{V}=\sum_\vp \frac{w_\vp}{\va_\vp-R_i}+\sum^N_{j=1}{}' \frac{2}{R_i-R_j}\, .\label{eg:R-Geq}
\ee
In this set of equations, the prime $(')$ excludes the $j=i$ term of the $j$ sum. $\va_\vp=\vp^2/2\mu_e$ denotes the free-pair kinetic energy, with $\mu_e=m_e/2$ being the electron pair reduced mass. Being valid for any $N$, these Richardson-Gaudin equations are quite appropriate for studying superconducting grains and nanostructures\cite{BraunPRL1998,DukelskyRMP,Gambacurta,Li2005,Ortiz2005,Garcia2005,Brihuega2011,MoniqEPJB2016} outside the thermodynamical limit.\

For conventional superconductors, the normal electron Fermi energy $\va_{F}$ is much larger than the potential extension width, $\Omega$, of the order of a phonon energy; so, the density of states in this layer is essentially constant. It has been shown\cite{CrouzeixPRL} that, for a pair density of states $\rho$ taken as constant, the ground-state energy of $N$ Cooper pairs in the thermodynamic limit takes a surprisingly simple form, within underextensive terms, as
\be
E_N=NE_1+\frac{N(N-1)}{2\rho}\frac{1+\sigma}{1-\sigma}\, ,\label{eq:ENCP}
\ee
where $\sigma=e^{-1/\rho V}$. (The factor 2 difference from the standard exponent comes from the fact that $\rho$ here is the pair density of states.) $E_1$ is the single-pair energy found by Cooper\cite{Cooper1956}, which can be obtained from Eq.~(\ref{eg:R-Geq}) taken for $N=1$, that is, without the second sum.

%\be
%\frac{1}{V}=\sum_\vp \frac{w_\vp}{\va_\vp-E_1}\, .\label{eq:singlepairsol}
%\ee

The exact cancellation in the $N$-Cooper pair energy of all extensive terms beyond $N(N-1)$ suggests that, under a BCS-like potential, a coboson only interacts with one among $N$ cobosons, as if it is completely ignorant of the other surrounding cobosons. The physics behind this astonishing many-body effect remains obscure up to now.\

Two reasons can lead to this strange result:

\noindent (i) The fact that the attractive potential given in Eq.~(\ref{def:V}) only acts  between zero-momentum pairs. To prove this first point, we must show that for a short-range potential between pairs having an {\it arbitrary} center-of-mass momentum, the $N$-coboson ground-state energy has extensive terms beyond $N(N-1)$.% We however wish to note that even though the potential form given in Eq.~(\ref{def:V}) proves to be necessary, it may not be sufficient to bring about such a magic cancellation.

\noindent (ii) The fact that the density of states in the energy extension where the potential acts is constant. To prove this second point and ascribe the mysterious cancellation to the system dimensionality, we must show that exact cancellation of extensive terms beyond $N(N-1)$ does not occur for other space dimension than 2D.\

 The present work is devoted to these two possible reasons. The understanding of the underlying physics that leads to the $N$-pair energy given in Eq.~(\ref{eq:ENCP}) will shed a new light on the merits and the limitations of the $V_{BCS}$ potential when used in the study of realistic systems.  \

To this end, we slightly modify the potential given in Eq.~(\ref{def:V}) in order for the present work to be applicable for cold-atom systems\cite{pethick2002,Bloch2008} which today are of major interest: the operators $(a^\dag, b^\dag)$ now denote the creation operators of the two fermion species at hand with reduced mass $\mu^{-1}=m_a^{-1}+m_b^{-1}$; the range of $\va_\vp=\vp^2/2\mu$ energy in which the attractive potential acts, that is, $w_\vp=1$, now extends from 0 to $\Omega$. Moreover, to study the first reason, we generalize the $V_{BCS}$ potential given in Eq.~(\ref{def:V}) into a short-range separable potential, $V_{CA}$, which acts between fermion pairs having an arbitrary center-of-mass momentum $\vK$
\be
V_{CA}=-V\sum_{\vK\vp\vp'} w_{\vp'}w_\vp B^\dag_{\vK\vp'}B_{\vK \vp}\,, \label{eq:VintofBKp}
\ee
where $B^\dag_{\vK\vp}\equiv a^\dag_{\vp+ \gamma_a\vK}b^\dag_{-\vp+\gamma_b \vK}$ with $\gamma_a=1-\gamma_b=m_a/(m_a+m_b)$. In translationally invariant systems as here considered, the center-of-mass momentum $\vK$ is conserved during particle scattering. \

The most striking result of this work is that for $N$ cobosons interacting via a BCS-like potential between zero-momentum pairs, the density expansion of their ground-state energy depends on space dimension $D$ in a very compact form for $D=(1,2,3)$, namely
\bea
\lefteqn{\frac{E_N{-}NE_1}{E_1}{=} {-}(N{-}1)(4{-}D)\bigg\{\nu_1(N\lambda^2)} \\
&&\!\! {+}(2{-}D)D\Big[\nu_2 (N\lambda^2)^2{+}\nu_3\!\left(24{-}14D{-}D^2\right)\!(N\lambda^2)^3{+}\cdots\!\Big]\!\bigg\}\, .\nn\label{enD123}
\eea
The $\nu_n$'s are numerical factors while $\lambda^2$ scales as $(a_B/L)^D$ with $L$ being the sample size and $a_B$ the single-pair Bohr radius; so, $N\lambda^2$ scales as the pair density $n$. This result confirms that, in 2D systems, the interaction part of the $N$-coboson energy is in $N(N-1)$ only. It further suggests that the whole interaction part should cancel for $D=4$, which would be even stranger because the system energy would then reduce to that of $N$ noninteracting cobosons. 4D systems actually are of physical interest because their density of states has the same energy dependence as 2D parabolic traps\cite{pethick2002}.  We will show in Sec.~\ref{sec:4Dtrap} that this seemingly pathological dimensionality-induced cancellation of all interaction terms does not occur in 4D. \

\section{Single-pair binding energy}
Let us first consider the ground-state binding energy of a single pair in $D=(1,2,3,4)$ dimension. Since a single pair in its ground state has zero center-of-mass momentum, the ground-state energy obtained for $V_{CA}$ coincides with that for $V_{BCS}$.

By turning discrete sum into integral, with a  density of states in $D$ dimension written as $\rho (\va/\Omega)^{(D-2)/2}$, where $\rho$ is the density of states at the upper potential cutoff $\Omega$, the single-pair binding energy follows from the $(-E_1)>0$ solution of Eq.~(\ref{eg:R-Geq}) taken for $N=1$, namely
\be
\frac{1}{V}=\sum_\vp \frac{w_\vp}{\va_\vp-E_1}\simeq\int_0^\Omega\rho d\va \frac{(\va/\Omega)^{(D-2)/2}}{\va-E_1}\, .\label{1VE_1solu}
\ee

\begin{figure}[t!]
\begin{center}
\includegraphics[trim=0.5cm 0cm 0.5cm 0.6cm,clip,width=3.3in] {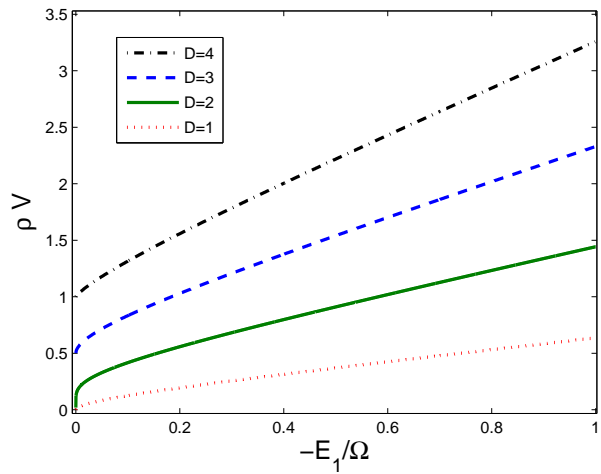}
   \caption{\small (color online) $\rho V$ as a function of $(-E_1)/\Omega$ for $D=(1,2,3,4)$, according to Eqs.~(\ref{rhoVD1}), (\ref{rhoVD2}), (\ref{rhoVD3}), and (\ref{rhoVD4}), respectively. }
   \label{Fig1}
\end{center}
\end{figure}

\noindent {\bf(i)} For $D=1$, the energy dependence of the density of states is in $1/\sqrt{\va}$; so,
\be
\frac{1}{\rho V}=\!\int^\Omega_0\!\!\frac{d\va}{\sqrt{\va/\Omega}} \frac{1}{ (\va-E^{(1)}_1)}=2\sqrt{\frac{\Omega}{-E^{(1)}_1}}\arctan\sqrt{\frac{\Omega}{-E^{(1)}_1}}\, .\label{rhoVD1}
\ee
This equation has a $(-E^{(1)}_1)>0$ solution for whatever $V$ (see Fig.~\ref{Fig1}). In the regime of physical interest, that is, $|E^{(1)}_1|\ll \Omega$, it reads
\be
E^{(1)}_1\simeq -\Omega (\rho V \pi)^2\, .
\ee

\noindent  {\bf(ii)} For $D=2$, the density of states is constant; so,
\be
\frac{1}{\rho V}=\int^\Omega_0d\va \frac{1}{ \va-E^{(2)}_1}=\ln \left(\frac{\Omega}{-E^{(2)}_1}+1\right)\, ,\label{rhoVD2}
\ee
from which we analytically derive the single-pair energy as
\be
E^{(2)}_1=-\Omega\frac{\sigma}{1-\sigma}
\ee
with $\sigma=e^{-1/\rho V}$ (see Fig.~\ref{Fig1}). This negative energy, which exists even for $V$ vanishingly small, corresponds to the binding energy of a single pair obtained by Cooper\cite{Cooper1956}: indeed, the effect of a large normal electron Fermi energy $\va_F$ is to make constant the density of states in the energy layer where the 3D potential acts.\

\noindent  {\bf(iii)} For $D=3$, the  energy dependence of the density of states is in $\sqrt{\va}$; so,
\be
\frac{1}{\rho V}=\int^\Omega_0 \!d\va \frac{ \sqrt{\va/\Omega}}{ \va-E^{(3)}_1}=2- 2\sqrt{\frac{-E^{(3)}_1}{\Omega}}\arctan  \sqrt{\frac{\Omega}{-E^{(3)}_1}}.\label{rhoVD3}
\ee
A solution $(-E^{(3)}_1)>0$ exists for $V$ larger than a threshold value $V^{(3)}_{th}=1/2\rho$ (see Fig.~\ref{Fig1}). In the regime of physical interest, that is, $|E^{(3)}_1|\ll \Omega$, this yields
\be
E^{(3)}_1\simeq -\frac{4}{\pi^2}\Omega\left(1-\frac{1}{2\rho V}\right)^2\, .
\ee

\noindent  {\bf(iv)} For $D=4$, the density of states is linear in $\va$; so, 
\bea
\frac{1}{\rho V}&=&\int^\Omega_0d\va \frac{\va/\Omega}{ \va-E^{(4)}_1}\nn\\
&=& 1 -\frac{-E^{(4)}_1}{\Omega}\ln \left(\frac{\Omega}{-E^{(4)}_1}+1\right)\label{rhoVD4}
\eea
has a $(-E^{(4)}_1)>0$ solution provided that $V$ is larger than the threshold value $V^{(4)}_{th}=1/\rho$. The threshold  for a single pair to form a bound state increases with space dimension, as seen in Fig.~\ref{Fig1}. In the physically relevant regime $|E^{(4)}_1|\ll \Omega$, the above equation yields
\be
E^{(4)}_1\simeq\Omega\frac{1-1/\rho V}{\ln (1-1/\rho V)-\ln(-\ln (1-1/\rho V))}\, .
\ee

\section{Density expansion parameter}

We are going to perform the density expansion of the energy of $N$ cobosons interacting via the $V_{CA}$ potential and the $V_{BCS}$ potential.\

In this study, the natural dimensionless parameter appears as
\be
\lambda^2\equiv \frac{2}{\rho \Omega }\left(\frac{\Omega}{-E_1}\right)^{D/2}\, .\label{eq:lambda2}
\ee
Since the density of states $\rho$ scales as the sample volume $L^D$, the product $N\lambda^2\propto n$ can be used as a parameter to study the small density expansion of the $N$-coboson energy. More precisely, $\lambda^2$ scales as $(a_B/L)^D$, where $a_B$ is the single-pair Bohr radius defined through $E_1\equiv -1/2\mu a_B^2$. To show it explicitly, we first note that the density of states in $D$ dimension is defined through
\be
\Theta_D\left(\frac{L}{2\pi}\right)^D  p^{D-1}dp=\rho \left(\frac{\va_\vp}{\Omega}\right)^{D/2-1}d\va_\vp\, ,
\ee
where  $\Theta_D=2\pi^{D/2}/\Gamma(D/2)$ is the  solid angle in $D$ dimension,  with $\Gamma(x)$ being the gamma function. As $\va_\vp=\vp^2/2\mu$, this gives
\be
\Theta_D\left(\frac{L}{2\pi}\right)^D =\frac{2\rho \Omega}{(2\mu \Omega)^{D/2}}\, .
\ee
So, for $E_1$ written in terms of $a_B$, Eq.~(\ref{eq:lambda2}) leads to
\be
\lambda^2=2\Gamma(D/2)\left(\frac{2\sqrt{\pi} a_B}{L}\right)^D\, .
\ee
This shows that the $N\lambda^2$ expansion we are going to perform, in fact, corresponds to an expansion in the usual dimensionless parameter  that controls many-body effects in the coboson systems\cite{MoniqPhysreport,book}, namely $\eta=N(a_B/L)^D$.  \

Another dimensionless coefficient that appears for cobosons interacting via the $V_{BCS}$ potential is
\be
\alpha_m=\frac{\lambda^2}{2}\sum_\vp w_\vp\frac{(-E_1)^{m+1}}{(\va_\vp-E_1)^{m+1}}=\int^{\frac{\Omega}{-E_1}}_0 dx \frac{x^{D/2-1}}{(x+1)^{m+1}}\, .\label{eq:alpha_n}
\ee
Since the integrand of the above equation decreases as $x^{D/2-2-m}$, we can safely extend the upper integral limit to infinity for $m\geq 1$ and $D=(1,2,3)$ in the physically relevant regime $|E_1|\ll \Omega$. This gives
\begin{subeqnarray}
\label{valuealpha1}
\alpha_1&=&\frac{\pi}{2} \quad{\rm for}\quad D=(1,3)\, ,\\
\alpha_1&=&1\quad {\rm for}\quad D=2\, .
\end{subeqnarray}
Higher $\alpha_m$'s are related through
\be
2m \alpha_m=(2m-D)\alpha_{m-1}\, ,\label{eq:alphanandn-1}
\ee
as obtained from Eq.~(\ref{eq:alpha_n}) by an integration by part. So, for $m\geq 2$,
\be
\alpha_m=\frac{(2m-D)\cdots (4-D)}{2^{m-1}m!}\alpha_1\, .\label{eq:alphanandn-2}
\ee
 As a result, all $\alpha_m$'s are finite. They form a geometrical series that depend on space dimension $D=(1,2,3)$, but not on the single-pair energy $E_1$ nor on the potential cutoff $\Omega$, provided that $|E_1|\ll \Omega$.

\section{Effect of arbitrary center-of-mass momentum}
We tackle the first question by considering cobosons having an arbitrary center-of-mass momentum $\vK$ and interacting through the short-range potential $V_{CA}$ given in Eq.~(\ref{eq:VintofBKp}). We want to study whether, in $D=(1,2,3)$ dimension, the density expansion of the $N$-coboson energy has extensive terms beyond $N(N-1)$. For this purpose, we consider the Hamiltonian mean value in the $N$-coboson state, namely
\be
\lan H\ran_N=\frac{\lan v|B_0^N  H B^{\dag N}_0|v\ran}{\lan v|B_0^N   B^{\dag N}_0|v\ran}\,,
\ee
where $B^\dag_0$ creates a ground-state coboson and $|v\ran$ is the vacuum state. $\lan H\ran_N$ corresponds to the Born value of the $N$-coboson ground-state energy. \

Using the coboson many-body formalism\cite{MoniqPhysreport,book}, we have shown that $\lan H\ran_N$ is given, within the present notations, by (see Eq.~(86) in Ref.~\onlinecite{Moniq2015PRA})
\bea
\lefteqn{\frac{\lan H\ran_N -NE_1}{E_1}\simeq}\label{eq:srsp1} \\
&&-(N-1)\Bigg[\frac{\alpha_2}{2\alpha^2_1}(N\lambda^2){+}\left(\frac{\alpha_2\alpha_3}{2\alpha_1^4}{-}\frac{\alpha_4}{4\alpha_1^3}\right)(N\lambda^2)^2{+}\cdots\Bigg],\nn
\eea
which, for $\alpha_n$'s given in Eq.~(\ref{eq:alphanandn-2}), reduces to
\bea
\lefteqn{\frac{\lan H\ran_N -NE_1}{E_1}\simeq-(N-1)(4-D) \bigg[\frac{1}{8\alpha_1}(N\lambda^2)}\label{eq:srsp2}\hspace{2cm} \\
&&+\frac{(6-D)(8-3D)}{768\alpha_1^2}(N\lambda^2)^2+\cdots\bigg].\nn
\eea
The above result evidences that for a short-range potential which acts between fermion pairs having an arbitrary center-of-mass momentum, the $N$-coboson ground-state energy in the Born approximation has terms beyond $N(N-1)$ in $D=(1,2,3)$ dimension --- but apparently not in 4D.

\section{Effect of space dimension}

We now turn to the second question related to the effect of space dimension on the ground-state energy of $N$ cobosons. To this end, we consider $N$ pairs having zero center-of-mass momentum and interacting via the $V_{BCS}$ potential given in Eq.~(\ref{def:V}) for $D=(1,2,3)$ dimension. Their energy follows from solving the Richardson-Gaudin equations given in Eq.~(\ref{eg:R-Geq}), for a density of states not necessarily constant.

\subsection{Resolution of Richardson-Gaudin equations}

To  solve these equations in an easy way, we rescale $R_i$ as $E_1(1- r_i)$ and expand $1/(\va_\vp-R_i)$ in powers of $r_i$ as
\be
\frac{1}{\va_\vp-R_i}=\sum_{m=0}^\infty \frac{(-E_1)^m}{(\va_\vp-E_1)^{m+1}}r_i^m\, .
\ee
The Richardson-Gaudin equations (\ref{eg:R-Geq}) then read, with the help of Eq.~(\ref{1VE_1solu}),
\be
0=\sum_{m=1}^\infty r_i^m \sum_\vp w_\vp\frac{(-E_1)^{m+1}}{(\va_\vp-E_1)^{m+1}} +{\sum_{j=1}^N}{}' \frac{2}{r_i-r_j}\label{eq:RGt_i}
\ee
that, using Eq.~(\ref{eq:alpha_n}), we rewrite as
\be
0=\sum_{m=1}^\infty \alpha_m r_i^m+\lambda^2{\sum_{j=1}^N}{}' \frac{1}{r_i-r_j}\, .\label{eq:RGt_i2}
\ee
To solve the above equation, we expand $r_i$ as
\be
r_i=a_i\lambda+b_i\lambda^2+c_i\lambda^3+d_i\lambda^4+e_i\lambda^5+f_i\lambda^6+\cdots\label{eq:t_iexpansion}
\ee
This expansion has been previously\cite{MoniqueEPJB2011} worked out up to $\lambda^4$ to obtain the cubic term in the density expansion of the $N$-Cooper pair ground-state energy, that is, for a constant density of states  made possible by the existence of a large normal electron Fermi energy.
We here wish not only to consider energy-dependent density of states in $D$ dimension, but also to go up to $\lambda^6$ to obtain the quartic term in density, in order to better control the effect of space dimension on the $N$-coboson correlation energy.\

To this end, we insert Eq.~(\ref{eq:t_iexpansion}) into Eq.~(\ref{eq:RGt_i2}), and expand $1/(r_i-r_j)$ in powers of $\lambda$. By matching the coefficients of the $\lambda$ and $\lambda^2$ terms, we get
\begin{subeqnarray}\label{eqs:coefficents}
0&=& \alpha_1 a_i+{\sum_{j}}' \frac{1}{a_i-a_j}\, ,\slabel{eqs:coefficents1}\\
0&=& \alpha_1  b_i +\alpha_2 a_i^2 -{\sum_{j}}' \frac{b_i-b_j}{(a_i-a_j)^2}\, .\slabel{eqs:coefficents2}
\end{subeqnarray}
Similar  but  heavier relations obtained from higher-order  terms in $\lambda$ are given in \ref{app:sec1}.
%\begin{subeqnarray}
%\beta_1&=& \frac{1}{a_{ij}}\, ,\\
%\beta_2&=& -\frac{b_{ij}}{a^2_{ij}}\, ,\\
%\beta_3&=& \frac{b^2_{ij}}{a^3_{ij}}-\frac{c_{ij}}{a^2_{ij}}\, ,\\
%\beta_4&=& -\frac{b^3_{ij}}{a^4_{ij}} +2\frac{b_{ij}c_{ij}}{a^3_{ij}}-\frac{d_{ij}}{a^2_{ij}}\, ,\\
%\beta_5&=& \frac{b^4_{ij}}{a^5_{ij}}-3\frac{b^2_{ij}c_{ij}}{a^4_{ij}} +2\frac{b_{ij}d_{ij}}{a^3_{ij}}+\frac{c^2_{ij}}{a^3_{ij}}-\frac{e_{ij}}{a^2_{ij}}\, .
%\end{subeqnarray}

The $N$-coboson ground-state energy
\be
E_N=\sum_{i=1}^N R_i= E_1(N -\mathcal{R}_N )
\ee
follows from $\mathcal{R}_N=\sum_ir_i$. It is possible to obtain its $\lambda$ expansion without having to determine the $r_i$'s individually. To do it, we first sum Eq.~(\ref{eq:RGt_i2}) over $i$. As the second sum reduces to zero by symmetry, this yields
\be
 0=\sum_{m=1}^\infty \alpha_m\sum_i r_i^m\, ,\label{eq:sumnalphanti}
\ee
in which we replace the $r_i$'s by their $\lambda$ expansion.\

(i) The $\lambda$ term immediately gives
\be
\sum_i a_i=0\, ,\label{sumai=0}
\ee
from which we conclude that $\mathcal{R}_N $ has no term in $\lambda$.\

(ii) The $\lambda^2$ term of Eq.~(\ref{eq:sumnalphanti}) gives
\be
0=\alpha_1\sum_i b_i+\alpha_2 \sum_i a_i^2\, .\label{eq:sumbiai2}
\ee
To obtain $\sum_i a_i^2$, we multiply Eq.~(\ref{eqs:coefficents1}) by $a_i$ and we sum over $i$. This yields
\be
0=\alpha_1 \sum_i a_i^2+\sum_i{\sum_{j}}' \frac{a_i}{a_i-a_j}\, .
\ee
As the double sum is equal to $(1/2)\sum_i \sum_{j}' (a_i-a_j)/(a_i-a_j)$, which readily gives $N(N-1)/2$, we get
\be
\sum_i a_i^2=-N(N-1)\frac{1}{ 2\alpha_1} \, .\label{eq:sumiai2}
\ee
Equation (\ref{eq:sumbiai2}) then gives the $\lambda^2$ term of $ \mathcal{R}_N $, with $\alpha_2$ obtained from Eq.~(\ref{eq:alphanandn-2}), as 
\be
\sum_i b_i=N(N-1)\frac{\alpha_2}{2\alpha_1^2}=\frac{N(N-1)}{8\alpha_1}(4-D)\, .\label{eq:condbi}
\ee

(iii) Similar calculations for the $\lambda^3$ and $\lambda^5$ terms of Eq.~(\ref{eq:sumnalphanti}), shown in \ref{app:sec2}, give 
\be
\sum_ic_i=0= \sum_ie_i\,,
\ee
which supports the fact that $\mathcal{R}_N$ has no odd term in $\lambda$. So, $\mathcal{R}_N$, hence $E_N$, has an analytical expansion in the coboson density,  which is rather reasonable.\

(iv) The $n^3$ term of $\mathcal{R}_N$ follows from the $\lambda^4$ term of $r_i$, that is, the sum of $d_i$'s. Using the $\alpha_m$'s given in Eq.~(\ref{eq:alphanandn-2}), we find their sum as (see \ref{app:sec2})
\bea
\sum_i d_i&\simeq&\frac{N^3}{2\alpha_1^4}\left(3\alpha_2\alpha_3-2\frac{\alpha_2^3}{\alpha_1}-\alpha_1\alpha_4\right)\nn\\
&=& \frac{N^3}{384\alpha_1^2}D(2-D)(4-D)\,.
\eea
%while for $D=4$, Eqs.~(\ref{eq:al14D}) and (\ref{eq:al24D}) give
%\be
%\sum_i d_i\simeq-\frac{N^3}{24 \alpha_1^3}\Big[\frac{3}{\alpha_1^2}-\frac{3}{\alpha_1}+1\Big]
%\ee

(v) In the same way, the $n^4$ term of $\mathcal{R}_N$ follows from the $\lambda^6$ term of $r_i$, that is, the sum of $f_i$'s. This sum is given (see \ref{app:sec2}) by
\bea
\sum_i f_i&\simeq&\frac{N^4}{8\alpha_1^4}\left(32\frac{\alpha_2^5}{\alpha_1^4}-92\frac{\alpha_2^3 \alpha_3}{\alpha_1^3}+48\frac{\alpha_2^2 \alpha_4}{\alpha_1^2}\right.\label{sumfi1}\\&&
\left.+45\frac{\alpha_2 \alpha_3^2}{\alpha_1^2}
-18\frac{\alpha_3 \alpha_4}{\alpha_1}-20\frac{\alpha_2 \alpha_5}{\alpha_1}+5\alpha_6\right)\, ,\nn
\eea
which also reads, with the help of Eq.~(\ref{eq:alphanandn-2}), in a surprisingly compact form as
\be
\sum_i f_i\simeq \frac{N^4}{36864\alpha_1^3}D(2-D)(4-D)\big(24-14D-D^2\big)\, .
\ee
%while for $D=4$, Eqs.~(\ref{eq:al14D}) and (\ref{eq:al24D}) give
%\be
%\sum_i f_i\simeq\frac{N^4}{192 \alpha_1^4}\Big(\frac{24}{\alpha_1^4}-\frac{46}{\alpha_1^3}+\frac{39}{\alpha_1^2}-\frac{18}{\alpha_1}+4\Big)\, .
%\ee

\subsection{$N$-coboson ground-state energy}
Combining the above results, we find that  in the thermodynamic limit, the expansion in density, $n\propto N\lambda^2$, of the $N$-coboson ground-state energy for $D=(1,2,3)$ dimension appears as given in Eq.~(\ref{enD123}), with $\nu_1=1/8\alpha_1$, $\nu_2=1/384 \alpha_1^2$, and $\nu_3=1/36864\alpha_1^3$
%\begin{widetext}
%\be
%\frac{E_N-NE_1}{NE_1}\simeq -(N-1)\lambda^2\left[ \frac{4-D}{8\alpha_1}+(4-D)(2-D)D\left(\frac{1}{384\alpha_1^2}N\lambda^2+\frac{24-14D-D^2}{36864\alpha_1^3}(N\lambda^2)^2+\cdots\right)\right] ,\label{GSenergyD123}
%\ee
%\end{widetext}
where $\alpha_1$ and $\lambda^2$ depend on space dimension $D$ according to Eq.~(\ref{valuealpha1}) and Eq.~(\ref{eq:lambda2}). This result confirms that for $D=2$, that is, for a constant density of states, the interaction part of the energy  for $N$ cobosons interacting via a BCS-like potential is in $N(N-1)$ only. By contrast, the exact cancellation of extensive terms beyond $N(N-1)$ does not occur for $D=(1,3)$. \

Moreover, from the above result, the whole interaction part of the ground-state energy seems to cancel for $D=4$, suggesting an even stranger many-body effect in 4D. Since the 4D density of states has the same energy dependence  as the one of 2D parabolic trap\cite{pethick2002}, 4D systems have  physical relevance. Actually, the previous calculations are not valid for 4D. On closer inspection, we found that the exact cancellation of all interaction terms does not occur in 4D, that is, in 2D parabolic trap, as we now show.

\section{4D or 2D parabolic trap\label{sec:4Dtrap}}

Let us first calculate the $\alpha_m$'s given in Eq.~(\ref{eq:alpha_n}) for $4D$. To get $\alpha_1$, we must keep the integral upper boundary $\Omega/(-E_1)$ in the integral to avoid logarithmic divergence.  We then find
\be
\alpha_1=\ln\left(1+\frac{\Omega}{-E_1}\right)-\frac{\Omega/(-E_1)}{1+\Omega/(-E_1)}  \,. \label{eq:al14D}
\ee
Contrary to $D=(1,2,3)$, the $\alpha_1$ coefficient depends on the single-pair energy $E_1$ and the cutoff $\Omega$  through their ratio. By contrast, $\alpha_m$'s for $m\geq 2$ can be obtained by extending the integral upper boundary to infinity. This gives
\be
\alpha_2= \frac{1}{2} \,.\label{eq:al24D}
\ee
Higher $\alpha_m$'s again form a geometric series, as in Eq.~(\ref{eq:alphanandn-1}), but the series now starts with $\alpha_2$. The fact that $\alpha_1$ does not belong to this geometric series makes the $D$-dependence of the interaction energy for $D=4$ different from  lower dimensions. The density expansion of the $N$-coboson ground-state energy   instead appears as
\bea
\lefteqn{\frac{E_N- NE_1}{E_1}\simeq}\nn\\
&& -(N-1)\left[ \frac{1}{4\alpha_1}(N\lambda^2)-\frac{\alpha_1^2-3\alpha_1+3}{24\alpha_1^4}(N \lambda^2)^2\right.\\
&&\left.+\frac{4\alpha_1^4-18\alpha_1^3+39\alpha_1^2-46\alpha_1+24}{192\alpha_1^8}(N\lambda^2)^3+\cdots\right] .\nn
\eea
So, in 4D or in 2D parabolic trap, the exact cancellation of all interaction terms in the $N$-coboson energy does not occur.\

Using these $\alpha_m$'s into Eq.~(\ref{eq:srsp1}), we also find that, for the above same reason, exact cancellation does not occur in the interaction energy of the Hamiltonian mean value for $N$ ground-state cobosons interacting via the $V_{CA}$ potential.

\section{Conclusion}

We study the ground-state energy of $N$ cobosons interacting through a BCS-like potential between zero-momentum fermion pairs, and through a short-range separable potential between fermion pairs having an arbitrary center-of-mass momentum. Our goal is to determine the effects of the potential characteristics and the space dimension on the interaction part of the $N$-coboson ground-state energy. We find that for 2D systems interacting via a BCS-like potential, the interaction part is in $N(N-1)$ only, in agreement with  previous results obtained in the context of Cooper pairs. Such a striking exact cancellation of the correlation energy exists uniquely for 2D systems. This mysterious cancellation results from the marriage of the very peculiar form of the reduced BCS potential which acts between zero-momentum pairs only, and the constant density of states where the potential acts. Our analysis based on the density expansion of the $N$-coboson energy shows that this cancellation does not occur for systems with energy-dependent density of states nor for cobosons having an arbitrary center-of-mass momentum. This microscopic understanding allows us to better appreciate the beauty of the reduced BCS potential originally proposed to understand standard superconductivity, and to be cautious of its possible limitations when used in other fields such as nuclear\cite{DukelskyRMP,Sandulescu,Dukelsky50} and cold-atom\cite{Bloch2008} physics.\

\section*{Acknowledgement}

M.C. acknowledges many fruitful visits to Academia Sinica and NCKU, Taiwan. S.-Y.S. acknowledges a three-month financial support from CNRS (France) as invited researcher at INSP in Paris. Y.-C.C. wishes to thank INSP for hospitality during his frequent visits to Paris. Work supported in part by Ministry of Science and Technology, Taiwan under contract MOST 104-2112-M-001-009-MY2.

\renewcommand{\thesection}{\mbox{Appendix~\Roman{section}}} %\section{Appendix}
\setcounter{section}{0}
\renewcommand{\theequation}{\mbox{A.\arabic{equation}}} %\section{Appendix}
\setcounter{equation}{0} %
\section{ \label{app:sec1}}
We here list the relations, obtained in the same way as Eq.~(\ref{eqs:coefficents}), for $\lambda^3$, $\lambda^4$, and $\lambda^5$ terms:
\begin{widetext}
\begin{subeqnarray}
0\!&=&\! \alpha_1  c_i +2\alpha_2 a_i b_i+ \alpha_3 a_i^3 + {\sum_{j}}' \left(\frac{(b_i-b_j)^2}{(a_i-a_j)^3}-\frac{c_i-c_j}{(a_i-a_j)^2}\right)\, \slabel{eqs:coefficents3},\\
0\!&=& \!\alpha_1  d_i +\alpha_2 (b_i^2 +2a_i c_i)+ 3\alpha_3  a_i^2b_i+\alpha_4  a_i^4 +{ \sum_{j}}' \left(-\frac{(b_i-b_j)^3}{(a_i-a_j)^4} +2\frac{(b_i-b_j)(c_i-c_j)}{(a_i-a_j)^3}-\frac{d_i-d_j}{(a_i-a_j)^2}\right)  \, ,\slabel{eqs:coefficents4} \\
0\!&=& \!\alpha_1  e_i +2\alpha_2 (a_id_i +b_i c_i)+ 3\alpha_3 (a_i^2c_i+a_i b_i^2)+4\alpha_4  a_i^3 b_i+\alpha_5  a_i^5 + {\sum_{j}}' \left(\frac{(b_i-b_j)^4}{(a_i-a_j)^5}-3\frac{(b_i-b_j)^2(c_i-c_j)}{(a_i-a_j)^4}\right. \nn\\
&&\left.+2\frac{(b_i-b_j)(d_i-d_j)}{(a_i-a_j)^3}+\frac{(c_i-c_j)^2}{(a_i-a_j)^3}-\frac{e_i-e_j}{(a_i-a_j)^2}\right)  \, .\slabel{eqs:coefficents5}
\end{subeqnarray}
\end{widetext}
These  equations are necessary to obtain the $\lambda$ expansion of $r_i$ up to $\lambda^6$.

\renewcommand{\theequation}{\mbox{B.\arabic{equation}}} %\section{Appendix}
\setcounter{equation}{0} %
\section{ \label{app:sec2}}

We here explicitly derive the $\lambda^3$, $\lambda^4$, $\lambda^5$, and $\lambda^6$ terms of $\mathcal{R}_N$. These straightforward but quite heavy calculations ultimately show that the $\mathcal{R}_N$ expansion only contains even powers of $\lambda$.

\subsection{$\lambda^3$ term}
The $\lambda^3$ term of $\mathcal{R}_N$ follows from $\sum_i c_i$.  The $\lambda^3$ coefficient in Eq.~(\ref{eq:sumnalphanti}) gives this sum through
\be
0= \alpha_1 \sum_i c_i +2\alpha_2 \sum_i a_i b_i+ \alpha_3 \sum_i a_i^3\, .\label{eq:condci}
\ee

$\bullet$ To calculate $\sum_i a_i^3$, we multiply Eq.~(\ref{eqs:coefficents1}) by $a^2_i$ and we sum over $i$. This gives
\be
0=\alpha_1 \sum_i a_i^3+\sum_i{\sum_{j}}' \frac{a^2_i}{a_i-a_j}\, .
\ee
The double sum gives zero since we can rewrite it as $(1/2)\sum_i {\sum_{j}}' (a^2_i-a^2_j)/(a_i-a_j)=(1/2)\sum_i {\sum_{j}}' (a_i+a_j)$ while $\sum_i a_i=0$, due to Eq.~(\ref{sumai=0}). So, $\sum_i a_i^3=0$.\

$\bullet$ To calculate $\sum_i a_i b_i$, we multiply Eq.~(\ref{eqs:coefficents2}) by $b_i$ and sum over $i$. This gives
\be
0= \alpha_1\sum_i a_i  b_i +\alpha_2\sum_i  a_i^3 -\sum_i {\sum_{j}}'a_i \frac{b_i-b_j}{(a_i-a_j)^2}\, .\label{eq:sumabia3}
\ee
As $\sum_i  a_i^3=0$, while the double sum also reads
\bea
-\sum_i {\sum_{j}}'a_i \frac{b_i-b_j}{(a_i-a_j)^2}=-\frac{1}{2}\sum_i {\sum_{j}}' \frac{b_i-b_j}{a_i-a_j}\nn\\
=-\sum_ib_{i} {\sum_{j}}' \frac{1}{a_i-a_j}=\alpha_1\sum_i a_i b_i\, ,
\eea
as obtained with the help of Eq.~(\ref{eqs:coefficents1}) for the $j$ sum, we conclude from Eq.~(\ref{eq:sumabia3}) that $\sum_i a_i  b_i=0$.\

$\bullet$ So, Eq.~(\ref{eq:condci}) ultimately gives
\be
\sum_i c_i=0\, .
\ee
As a result,  the $\lambda^3$ term of $\mathcal{R}_N$ is equal to zero.\

\subsection{$\lambda^4$ term}
The $\lambda^4$ term of $\mathcal{R}_N$ follows from $\sum_i d_i$. The $\lambda^4$ coefficient of Eq.~(\ref{eq:sumnalphanti}) gives this sum through
\bea
0&=& \alpha_1 \sum_i d_i +\alpha_2 \sum_i (b_i^2 +2a_i c_i)+ 3\alpha_3 \sum_i a_i^2b_i\nn\\
&&+\alpha_4 \sum_i a_i^4\, .\label{eq:conddi}
\eea
$\bullet$ To calculate $\sum_i a_i^4$, we multiply Eq.~(\ref{eqs:coefficents1}) by $a^3_i$ and we sum over $i$. This gives
\be
0=\alpha_1 \sum_i a_i^4+\sum_i{\sum_{j}}' \frac{a^3_i}{a_i-a_j}\, .
\ee
We rewrite the double sum in the above equation as
\bea
\sum_i{\sum_{j}}' \frac{a^3_i}{a_i-a_j}&=&\frac{1}{2}\sum_i{\sum_{j}}' \frac{a^3_i-a_j^3}{a_i-a_j}\nn\\
&=&\frac{1}{2}\sum_i{\sum_{j}}' (a^2_i+a_ia_j+a_j^2)\, ,
\eea
which gives $(N-3/2)\sum_i a_i^2 +(1/2)(\sum_i a_i)^2$. Using Eq.~(\ref{eq:sumiai2}) and $\sum_i a_i=0$, we end up with
\be
\sum_i a_i^4= N(N-1)(2N-3)\frac{1}{4\alpha_1^2}\, .\label{eq:sumiai4}
\ee

$\bullet$ To calculate $\sum_i a_i^2b_i$, we multiply Eq.~(\ref{eqs:coefficents2}) by $a^2_i$ and we sum over $i$. This  gives
\be
0=\alpha_1 \sum_i a_i^2 b_i+ \alpha_2\sum_i a_i^4-\sum_i{\sum_{j}}' \frac{a^2_i (b_i-b_j)}{(a_i-a_j)^2}\, .\label{eq:sumiai2biai41}
\ee
We rewrite the double sum in the above equation as
\bea
\lefteqn{-\sum_i{\sum_{j}}' \frac{a^2_i (b_i-b_j)}{(a_i-a_j)^2}=-\frac{1}{2}\sum_i{\sum_{j}}' \frac{(a^2_i-a_j^2) (b_i-b_j)}{(a_i-a_j)^2}}\hspace{7cm}\nn\\
=-\frac{1}{2}\sum_i{\sum_{j}}' \frac{(a_i+a_j) (b_i-b_j)}{a_i-a_j}\, .
\eea
Since $\sum_i{\sum_{j}}' (b_i+b_j) (a_i+a_j)/(a_i-a_j)=0$, while $(a_i+a_j)/(a_i-a_j)=(-1+2a_i)/(a_i-a_j)$, the RHS of the above equation also reads
\be
-\sum_i  b_{i}{\sum_{j}}'\left(-1+ \frac{2a_i}{a_i-a_j}\right)\, ,
\ee
which, with the help of Eq.~(\ref{eqs:coefficents1}), gives $(N-1)\sum_i b_i +2\alpha_1 \sum_i a_i^2b_i$. Combining this result with Eqs.~(\ref{eq:condbi}) and (\ref{eq:sumiai4}), we end up with
\be
 \sum_i a_i^2 b_i=-N(N-1)(4N-5)\frac{\alpha_2}{12\alpha_1^3}\, .\label{eq:3sumiai2bi}
\ee

$\bullet$ To calculate the second term of Eq.~(\ref{eq:conddi}), we multiply Eq.~(\ref{eqs:coefficents3}) by $a_i$ and we sum over $i$. This  gives
\bea
0&=&\alpha_1 \sum_i a_i c_i+ 2\alpha_2\sum_i a_i^2 b_i+\alpha_3 \sum_i a_i^4\label{eq:sumiai2biai4}\\
&& +\sum_i{\sum_{j}}'a_i\left( \frac{(b_i-b_j)^2}{(a_i-a_j)^3}-\frac{c_i-c_j}{(a_i-a_j)^2}  \right)\, .\nn
\eea
The double sum in the above equation also reads
\be
\frac{1}{2}\sum_i{\sum_{j}}'(a_i-a_j)\left( \frac{(b_i-b_j)^2}{(a_i-a_j)^3}-\frac{c_i-c_j}{(a_i-a_j)^2}\right) ,
\ee
which reduces to
\be
\sum_i b_i{\sum_{j}}'\frac{b_i-b_j}{(a_i-a_j)^2}-\sum_ic_i{\sum_{j}}' \frac{1}{a_i-a_j}\, .
\ee
Using Eq.~(\ref{eqs:coefficents}) for the two sums over $j$, we end up with
\be
0=\alpha_1\sum_i( b_i^2+2a_i c_i)+ 3\alpha_2 \sum_i a_i^2 b_i+\alpha_3 \sum_i a_i^4\, .\label{eq:sumbi22aici}
\ee

$\bullet$ By combining Eqs.~(\ref{eq:conddi}), (\ref{eq:sumbi22aici}), (\ref{eq:sumiai4}), and (\ref{eq:3sumiai2bi}), we get the prefactor of the $\lambda^4$ term in $\mathcal{R}_N$ as
\bea
\sum_i d_i&=&\frac{N(N-1)}{4\alpha_1^4}\Big[(6N-8)\alpha_2\alpha_3-(4N-5)\frac{\alpha_2^3}{\alpha_1}\nn\\
&&-(2N-3)\alpha_1\alpha_4\Big]\, .
\eea

\subsection{$\lambda^5$ term}
The $\lambda^5$ term of $\mathcal{R}_N$ follows from $\sum_ie_i$. The $\lambda^5$ coefficient of Eq.~(\ref{eq:sumnalphanti}) gives this sum through
\bea
0&=& \alpha_1 \sum_i e_i +2\alpha_2 \sum_i (a_id_i +b_i c_i)\label{eq:condei}\\
&&+ 3\alpha_3 \sum_i (a_i^2c_i+a_i b_i^2)+4\alpha_4 \sum_i a_i^3 b_i+\alpha_5 \sum_i a_i^5\, .\nn
\eea

$\bullet$ To calculate $\sum_i a_i^5$, we multiply Eq.~(\ref{eqs:coefficents1}) by $a^4_i$ and sum over $i$. This  gives
\be
0=\alpha_1 \sum_i a_i^5+\sum_i{\sum_{j}}' \frac{a^4_i}{a_i-a_j}\, .
\ee
We rewrite the double sum in the above equation as
\bea
\sum_i{\sum_{j}}' \frac{a^4_i}{a_i-a_j}&=&\frac{1}{2}\sum_i{\sum_{j}}' \frac{a^4_i-a_j^4}{a_i-a_j}\\
&=&\frac{1}{2}\sum_i{\sum_{j}}' (a^3_i+a^2_ia_j+a_ia_j^2+a_j^3)\, ,\nn
\eea
which leads to $(N-2)\sum_i a_i^3 +\sum_i a_i^2\sum_i a_i$; so, $\sum_i a_i^5=0$, since $\sum_ia_i=0=\sum_i a_i^3$. \

$\bullet$ To calculate $\sum_i a_i^3 b_i$, we multiply Eq.~(\ref{eqs:coefficents2}) by $a^3_i$ and we sum over $i$. This  gives
\be
0=\alpha_1 \sum_i a_i^3 b_i+\sum_i a_i^5-\sum_i{\sum_{j}}' \frac{a^3_i(b_i-b_j)}{(a_i-a_j)^2}\, .\label{eq:ai3biai5}
\ee
The double sum in the above equation also reads
\bea
-\frac{1}{2}\sum_i{\sum_{j}}' \frac{(a^3_i-a_j^3)(b_i-b_j)}{(a_i-a_j)^2}\nn\\
=-\sum_i b_i{\sum_{j}}' \frac{a^2_i+a_ia_j+a_j^2}{a_i-a_j}\, .
\eea
By writing $a^2_i+a_ia_j+a_j^2$ as $(a_i-a_j)^2+3a_ia_j$ and $a_j/(a_i-a_j)$ as $-1+a_i/(a_i-a_j)$, the above term is equal to
\be
-\sum_i b_i{\sum_{j}}'(a_i-a_j)+3 \sum_i a_i b_i{\sum_{j}}'1-3\sum_ia_i^2 b_i{\sum_{j}}'\frac{1}{a_i-a_j}\, .
\ee
The sum of the first two terms give $(2N-3)\sum_ia_ib_i+\sum_ia_i \sum_i b_i=0$, while using Eq.~(\ref{eqs:coefficents1}), the last term gives $3\alpha_1\sum_ia_i^3 b_i$. Since $\sum_ia_i^5=0$, we find from Eq.~(\ref{eq:ai3biai5}) that $\sum_ia_i^3 b_i=0$.\

$\bullet$ To calculate the third term of Eq.~(\ref{eq:condei}), we multiply Eq.~(\ref{eqs:coefficents3}) by $a^2_i$ and we sum over $i$. This gives
\bea
0&=&\alpha_1 \sum_i a_i^2 c_i+ 2\alpha_2\sum_i a_i^3b_i+\alpha_3\sum_i a_i^5\label{eq:ai2ciai3bi}\\
&&+\sum_i{\sum_{j}}'a_i^2\left( \frac{(b_i-b_j)^2}{(a_i-a_j)^3}-\frac{c_i-c_j}{(a_i-a_j)^2}\right)\, .\nn
\eea
The double sum also reads
\be
\frac{1}{2}\sum_i{\sum_{j}}'(a_i^2-a_j^2)\left( \frac{(b_i-b_j)^2}{(a_i-a_j)^3}-\frac{c_i-c_j}{(a_i-a_j)^2}\right)\, ,
\ee
which is equal to
\be
\sum_i b_i{\sum_{j}}'(a_i+a_j)\frac{b_i-b_j}{(a_i-a_j)^2}- \sum_i c_i{\sum_{j}}'\frac{a_i+a_j}{a_i-a_j}\, .\nn
\ee
B writing $(a_i+a_j)/(a_i-a_j)$ as $-1+2a_i/(a_i-a_j)$, we find from Eq.~(\ref{eq:ai2ciai3bi}) and (\ref{eqs:coefficents}) that
\bea
0&=&3\alpha_1 \sum_i (a_i^2 c_i+a_ib_i^2)+ 4\alpha_2\sum_i a_i^3b_i+\alpha_3\sum_i a_i^5\nn\\
&&+(N-1)\sum_i c_i\, .
\eea
Since the last three terms are equal to zero, we end up with $\sum_i (a_i^2 c_i+a_ib_i^2)=0$. \

$\bullet$ To calculate the second term, we multiply Eq.~(\ref{eqs:coefficents4}) by $a_i$ and we sum over $i$. Similar calculation for the double sum leads to
%\bea
%0&=& \alpha_1 \sum_ia_i d_i +\alpha_2 \sum_i(a_ib_i^2 +2a^2_i c_i)+ 3\alpha_3 \sum_i a_i^3b_i\nn\\
%&&+\alpha_4 \sum_i a_i^5 +\sum_i {\sum_{j}}' a_i\left(-\frac{b^3_{ij}}{a^4_{ij}} +2\frac{b_{ij}c_{ij}}{a^3_{ij}}-\frac{d_{ij}}{a^2_{ij}}\right)\nn\\
%\eea
\bea
0&=&2\alpha_1 \sum_i(a_i d_i+b_ic_i) +3\alpha_2 \sum_i(a^2_i c_i+a_ib_i^2)\nn\\
&&+ 4\alpha_3 \sum_i a_i^3b_i+\alpha_4 \sum_i a_i^5\, .
\eea
Since the last three terms are equal to zero, we end up with $\sum_i(a_i d_i+b_ic_i)=0$.

$\bullet$ Equation (\ref{eq:condei}) then gives
\be
\sum_i e_i=0\,.
\ee
So, $\mathcal{R}_N$ has no term in $\lambda^5$.

\subsection{$\lambda^6$ term}

The $\lambda^6$ term of $\mathcal{R}_N$ follows from $\sum_i f_i$.  The $\lambda^6$ coefficient in Eq.~(\ref{eq:sumnalphanti}) gives this sum through
\bea
0&=& \alpha_1 \sum_i f_i +\alpha_2 \sum_i (c_i^2+2a_ie_i +2b_i d_i)\nn\\
&&+ \alpha_3 \sum_i (b_i^3+3a_i^2d_i+6a_i b_i c_i)\label{eq:condfi}
\\
&&+\alpha_4 \sum_i (4 a_i^3c_i+6a_i^2 b_i^2)+5\alpha_5 \sum_i a_i^4 b_i+\alpha_6 \sum_i a_i^6.\nn
\eea

$\bullet$ To calculate $\sum_i a_i^6$, we multiply Eq.~(\ref{eqs:coefficents1}) by $a_i^5$ and we sum over $i$. Since $\sum_i a_i=0=\sum_i a_i^3$, we get
\be
0=\alpha_1\sum_i a_i^6+\frac{2N-5}{2}\sum_i a_i^4+\frac{1}{2}\Big(\sum_i a_i^2\Big)^2\, .\label{eq:a6}
\ee

$\bullet$ To calculate $\sum_i a_i^4 b_i$, we  multiply Eq.~(\ref{eqs:coefficents2}) by $a_i^4$ and we sum over $i$. Similar algebras lead to
\bea
0&=&5 \alpha_1 \sum_i a_i^4 b_i+\alpha_2 \sum_i a_i^6+3(N-2)\sum_a a_i^2b_i\nn\\
&&+\Big(\sum_ib_i\Big) \Big(\sum_i a_i^2\Big)\, .
\eea

$\bullet$ To calculate the fourth term of Eq.~(\ref{eq:condfi}), we multiply Eq.~(\ref{eqs:coefficents3}) by $a_i^3$ and we sum over $i$. This yields
\bea
0&=&\alpha_1 \sum_i (4 a_i^3c_i+6a_i^2 b_i^2)+5\alpha_2\sum_i a_i^4 b_i+\alpha_3\sum_i a_i^6\nn\\
&&+\frac{2N-3}{2}\sum_i(b_i^2+2a_ic_i)+\frac{1}{2}\Big(\sum_ib_i\Big)^2\, .
\eea

$\bullet$ To calculate the third term, we multiply Eq.~(\ref{eqs:coefficents4}) by $a_i^2$ and we sum over $i$. This yields
\bea
0&=&\alpha_1 \sum_i(b_i^3+3a_i^2d_i+6a_i b_i c_i)+\alpha_2 \sum_i (4 a_i^3c_i+6a_i^2 b_i^2)\nn\\
&&+5\alpha_3\sum_i a_i^4 b_i+\alpha_4\sum_i a_i^6+(N-1)\sum_i d_i\, .
\eea

$\bullet$ To calculate the second term of Eq.~(\ref{eq:condfi}), we multiply Eq.~(\ref{eqs:coefficents4}) by $b_i$ and Eq.~(\ref{eqs:coefficents5}) by $a_i$ and we sum over $i$. By adding these two equations, we get
\bea
0=\!\alpha_1\!\! \sum_i(c_i^2+2a_ie_i+2 b_i d_i)+\alpha_2 \!\!\sum_i(b_i^3+3a_i^2d_i+6a_i b_i c_i)\nn\\
+\alpha_3\!\! \sum_i (4 a_i^3c_i+6a_i^2 b_i^2)+5\alpha_4\!\!\sum_i a_i^4 b_i+\alpha_5\!\!\sum_i a_i^6.\label{eq:c2aebd}\hspace{1cm}
\eea

Equations (\ref{eq:condfi}-\ref{eq:c2aebd}) ultimately lead to the sum of $f_i$'s given in Eq.~(\ref{sumfi1}).
%\bea
%\sum_i f_i&\simeq&\frac{N^4}{8\alpha_1^4}\left(32\frac{\alpha_2^5}{\alpha_1^4}-92\frac{\alpha_2^3 \alpha_3}{\alpha_1^3}+48\frac{\alpha_2^2 %\alpha_4}{\alpha_1^2}\right.\\&&
%\left.+45\frac{\alpha_2 \alpha_3^2}{\alpha_1^2}
%-18\frac{\alpha_3 \alpha_4}{\alpha_1}-20\frac{\alpha_2 \alpha_5}{\alpha_1}+5\alpha_6\right)\, .\nn
%\eea

\end{document}